\documentclass[twocolumn,showpacs,preprintnumbers,amsmath,amssymb]{revtex4}
\usepackage{tipa}
\usepackage{amssymb}
\usepackage{stmaryrd}
\usepackage{txfonts}
\usepackage{color}
\usepackage{subfigure}
\usepackage{graphicx}
\usepackage{dcolumn}
\usepackage{bm}
\usepackage{float}
\begin{document}

\preprint{}
\title{Giant photonic spin Hall effect near the Dirac points}
\author{Wenhao Xu$^{1}$}
\author{Qiang Yang$^{1}$}
\author{Guangzhou Ye$^{1}$}
\author{Weijie Wu$^{1}$}
\author{Wenshuai Zhang$^{1}$}
\author{Hailu Luo$^{1,2}$}\email{hailuluo@hnu.edu.cn}
\author{Shuangchun Wen$^{2}$}
\address{$^{1}$ Laboratory for Spin Photonics, School of
 Physics and Electronics, Hunan University, Changsha 410082, China\\
$^{2}$ Key Laboratory for Micro-/Nano-Optoelectronic Devices of Ministry of Education, School of
 Physics and Electronics, Hunan University, Changsha 410082, China}
\date{\today}

\begin{abstract}
The origin of spin-orbit interaction of light at a conventional optical interface lies in the transverse nature of the photon
polarization: The polarizations associated with
the plane-wave components experience slightly different
rotations in order to satisfy the transversality after reflection or refraction.
Recent advances in topological photonic materials provide crucial opportunities to reexamine the spin-orbit interaction of light at the unique optical interface.
Here, we establish a general model to describe the spin-orbit interaction of light in the photonic Dirac metacrystal. We find a giant photonic spin Hall effect near the Dirac points when a Gaussian beam impinges at the interface of the photonic Dirac metacrystal.
The giant photonic spin Hall effect is attribute to the strong spin-orbit interaction of light, which manifests itself as the large polarization rotations of different plane-wave components. We believe that these results may provide insight
into the fundamental properties of the spin-orbit interaction of
light in the topological photonic systems.
\end{abstract}
\pacs{42.25.-p, 42.79.-e, 41.20.Jb}

\keywords{spin-orbit interaction, photonic spin Hall effect, Dirac point}
\maketitle

\section{Introduction}\label{SecI}
 In classic electrodynamics, the reflection of plane wave at a dielectric interface can be determined directly from geometric optics at macroscopic scales~\cite{Jackson1999}, whereas for a real optical beam is not entirely appropriate. It has been pointed out that the wave evolution does not strictly governed by the traditional optics picture due to spin-orbit interactions~\cite{Onoda2004,Hosten2008,Bliokh2015}. The photonic spin Hall effect(SHE) manifesting itself as spin-dependent splitting in the light-matter interaction is considered as a result of the spin-orbit interaction of light~\cite{Bliokh201502,yinxb,Ling2017,Korger2014}, which plays a key role in many fields, including quantum weak measurement~\cite{zhoux2012,chens2017,ChenS2018}, two-dimensional material characterization ~\cite{zhouxpra2012,qiu2014}, mapping of absorption mechanisms~\cite{M¨¦nard2010}, and identification of multilayer graphene~\cite{zhoux2012,Kamp2016,cai2017}.

Topological metamaterials, including Chern insulators, topological insulators, Weyl semimetals and Dirac semimetals represent a unique effective medium approach for studying topological behaviors of electromagnetic waves, and have attracted growing research interest in various field. Recently there has been realization of topological insulators~\cite{ChenWJ2014,HeC2016,Slobo2017}, Weyl degeneracies~\cite{XuSY2015,YangB2017,XiaoM2017}, and Dirac degeneracies~\cite{LiuZK2014,SOL2016,GuoQ2017} in the metamaterial and metacrystal systems.
Thus the emergence of photonic topological metacrystal render a important foundation to investigate photonic SHE at the unique optical interface. Unlike graphene~\cite{Cast2009,Wu2017,Kamp2018} and silene~\cite{Wu2018}, 3D photonic topological metacrystal are nonplanar and and possess intrinsic spin-orbit coupling that results in unique band structure. These structures, especially for Dirac points and Weyl points, exhibit unusual physical properties that cannot be found in either monolayers or in bulk materials. In photonics , the fourfold degeneracy of Dirac point usually realized through various space symmetries and exhibit a bulk Hall effect due to time-reversal symmetry breaking~\cite{Slobo2017,Lul2016}. However, the conventional model can not present an accurate description of photonic SHE in the photonic  Dirac metacrystals. A fundamental understanding of the photonic SHE near the Dirac point in photonic topological materials is therefore essential for profound insight of the Dirac materials and even other photonic topological systems.

In this paper, we establish a general model to describe the behaviors of light at the interface of three dimensional metacrystal. In our model, the photonic SHE near the Dirac point in reflected field are taken into accounted. Both the transverse and in plane spatial shifts are obtained when a light beam impinges on the surface of metacrystals. We found that the strong photonic SHE occurs near the Dirac point, which are sensitive to incident angle, frequency, and the electromagnetic properties of material. Hence, we demonstrate the giant photonic SHE with various incident condition and give the corresponding large polarization rotation state as an explanation. Our result will enable us to better understand the spin-orbit interaction of light and will provide insights into the optical measurement of Dirac points, Weyl points, and exceptional points.

\begin{figure*}
\centerline{\includegraphics[width=18cm]{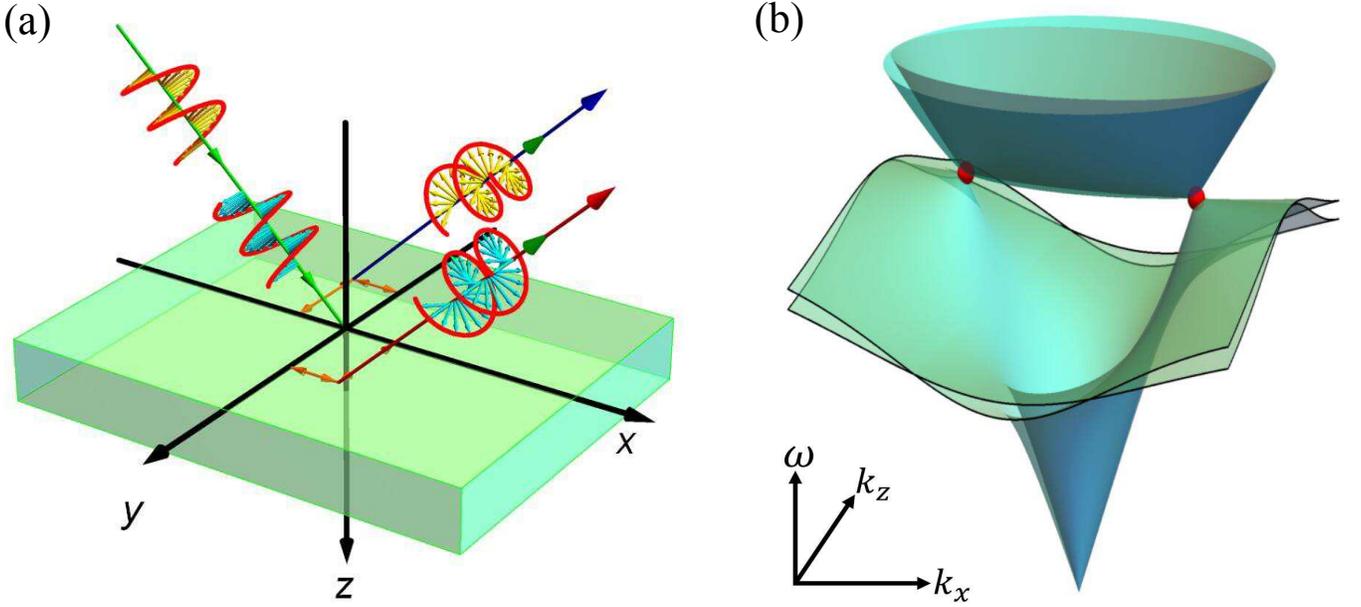}}
\caption{\label{Fig1} (a) Schematic illustrating the wave reflection near the Dirac point at the interface of Dirac metacrystal in a Cartesian coordinate system. On the interface of 3D metacrystal, the photonic SHE occurs which manifests as the spin-dependent splitting. (b) The effective bulk band structure of metacrystal on wave vector plane . Two bands are nearly overlapping with each other and there are fourfold band degeneracies at two points (marked by the red spheres) which are symmetrically placed.
}
\end{figure*}

\section{A general model for spin-orbit interaction of light}\label{SecII}
Here we first present photonic Dirac points in a medium with homogeneous effective electromagnetic properties. In our case, the general permittivity and permeability tensors of medium have the anisotropic form as $\bm\varepsilon=\varepsilon_{0}(\varepsilon_{x}\hat{x}\hat{x}+\varepsilon_{y}\hat{y}\hat{y}
+\varepsilon_{z}\hat{z}\hat{z})$ and $\bm\mu=\mu_{0}(\mu_{x}\hat{x}\hat{x}+\mu_{y}\hat{y}\hat{y}
+\mu_{z}\hat{z}\hat{z})$, respectively.
Then we assume $\varepsilon_{z}=\varepsilon_{y}$, $\mu_{z}=\mu_{y}$ are constants. Note that the magneto-electric tensors for the proposed metacrystal are not considered here. The presence of resonance along the $x$ direction for both permittivity and permeability indicates that there exist two bulk plasmon modes\cite{GuoQ2017}, a longitudinal electric mode and a longitudinal magnetic mode,
\begin{equation}       
\varepsilon_{x}=1+f_{1}\omega_{0}^2/(\omega_{0}^2-\omega^{2}),\qquad   \mu_{x}=1+f_{2}\omega^2/(\omega_{0}^2-\omega^{2}), \label{eq2}
\end{equation}
where $\omega_0$ indicates the resonance frequency and coefficients $f_1$ and $f_2$ are constants determined by the structure parameters which are adjustable.

Let us consider that a Gaussian beam of frequency $\omega$ impinges at an angle $\theta_{i}$ at the interface between vacuum and the metacrystal. $k_{i}=\omega /\sqrt{\varepsilon_{0}\mu_{0}}=\omega/c$ is the incident vector, and $c$ is the speed of light, and we assume the incident wave vector lies on the $\bm{xz}$ plane, as shown in Fig. 1(a). The band structure for a certain set of parameters satisfying the condition are shown in Fig.1(b), the location of Dirac point can obtain from the eigenvalue of Hamiltonian formalism in the propose metacrystals\cite{GuoQ2017}, $ k_{x}=k_{i}\sin\theta_{i}=\pm k_{i}\sqrt{\varepsilon_{y}\mu_{y}}$ ~~at the critical frequency, $\omega_{D}=\sqrt{1+f_{1}}\omega_{0}$.

Considering the optical axis direction, we introduced a unit vector $\mathcal{I}$, and let $\alpha,\beta,\gamma$ be direction cosines of the optical axis relative the cartesian laboratory frame~\cite{Lekner1991}:
\begin{equation}       
\mathcal{I}=\alpha~\hat{\textbf{x}}+\beta~\hat{\textbf{y}}+\gamma~\hat{\textbf{z}},
\end{equation}
where $\hat{\textbf{x}}$, $\hat{\textbf{y}}$ and $\hat{\textbf{z}}$ are unit vectors along the $x$, $y$ and $z$ positive axes. Since $\mathcal{I}$ is also a unit vector, thus $\alpha^{2}+\beta^{2}+\gamma^{2}=1$.

With $\Delta\varepsilon=\varepsilon_{x}-\varepsilon_{y}$ and $\Delta\mu=\mu_{x}-\mu_{y}$, the tensor reduces to \cite{Lekner1991}:

\begin{eqnarray}       
&\bm{\varepsilon}=
\left|                 
  \begin{array}{ccc}   
   \varepsilon_{y}+\alpha^2\Delta\varepsilon & \alpha\beta\Delta\varepsilon & \alpha\gamma\Delta\varepsilon \\  
    \alpha\beta\Delta\varepsilon &  \varepsilon_{y}+\beta^2\Delta\varepsilon & \beta\gamma\Delta\varepsilon  \\  
    \alpha\gamma\Delta\varepsilon & \beta\gamma\Delta\varepsilon &  \varepsilon_{y}+\gamma^2\Delta\varepsilon  \\  
  \end{array}
~\right|,
\\
&\bm{\mu}=
\left|                 
  \begin{array}{ccc}   
   \mu_{y}+\alpha^2\Delta\mu & \alpha\beta\Delta\mu & \alpha\gamma\Delta\mu \\  
    \alpha\beta\Delta\mu &  \mu_{y}+\beta^2\Delta\mu & \beta\gamma\Delta\mu  \\  
    \alpha\gamma\Delta\mu & \beta\gamma\Delta\mu &  \mu_{y}+\gamma^2\Delta\mu  \\  
  \end{array}
\right|.
\end{eqnarray}

After applying the constitutive relations and Maxwell's equation, we can obtain the characteristic equation. There are two normal (to the interface) wave vectors components $\bm{k}=({k_x},0,k_{z}^{\pm}) $ which are associated with eigenwaves determined by the nullspace of Hamiltonian formalism, given as $\bm{E^{\pm}} = {E_{0}} \bm{(e_{x}^{\pm},e_{y}^{\pm},e_{z}^{\pm})}$, where $E_{0}$ are the electric field magnitudes for the plus and minus modes, respectively. Likewise the corresponding magnetic fields are also obtained by constitutive relation, $\bm{H^{\pm}}=\frac{1}{\omega}\mu^{-1}k\times\bm{E^{\pm}}=\frac{E_{0}}{\eta_{0}}\bm{(h_{x}^{\pm},h_{y}^{\pm},h_{z}^{\pm})}$, where $\eta_{0}=\sqrt{\mu_{0}/\varepsilon_{0}}$.

To obtain a more accurate model, the arbitrary wave vector of Gaussian beam should be take into account. Here we noted that $k_{ix}$ and $k_{iy}$ respects the $x$ and $y$ components of arbitrary vector relative to beam center, respectively. For the sake of simplicity, we set the optical axis almost along the $\bm{x}$ axis, $\alpha\rightarrow 1$ and other components $\beta$ and $\gamma$ can be indicated the arbitrary wave vector $k_{ix}$ and $k_{iy}$, $\beta\sim\frac{k_{iy}}{k_{i}}, \gamma\sim\frac{k_{ix}}{k_{i}}\cos{\theta_{i}}$. Considering the propagation of a Gaussian beam under the first order paraxial approximation, the expressions of two vectors $k_{z}^{\pm}$ are rewritten as $q_e$ and $q_o$:

\begin{eqnarray}
\begin{aligned}
&&{q_{e}}=\sqrt{\dfrac{(-\sin^{2}\theta_{i}+\varepsilon_{y}\mu_{y})\varepsilon_{x}}{\varepsilon_{y}}}, \\
&&{q_{o}}=\sqrt{\dfrac{(-\sin^{2}\theta_{i}+\varepsilon_{y}\mu_{y})\mu_{x}}{\mu_{y}}}.
\end{aligned}
\end{eqnarray}

From boundary conditions~\cite{ChenRL2014} and paraxial approximation, the Fresnel¡¯s coefficients are obtained as:

\begin{eqnarray}
&r_{pp}=\dfrac{{q_{e}}-\varepsilon_{y}\cos\theta_{i}}{{q_{e}}+\varepsilon_{y}\cos\theta_{i}},
\nonumber  \\
&r_{ss}=\dfrac{-{q_{o}}+\cos\theta_{i}}{{q_{o}}+\cos\theta_{i}},
\label{Rco1}
\end{eqnarray}

\begin{eqnarray}
&r_{ps}=\dfrac{2~\gamma~\sqrt{\varepsilon_{x}\mu_{x}}(\sqrt{\varepsilon_{y}\mu_{y}}-\sqrt{\varepsilon_{x}\mu_{x}})\cos\theta_{i}}{({q_{e}}+\varepsilon_{x}\cos\theta_{i})({{q_{o}}+\mu_{x}\cos\theta_{i}})},
\nonumber  \\
&r_{sp}=\dfrac{2~\beta~\sqrt{\varepsilon_{x}\mu_{x}}(\sqrt{\varepsilon_{y}\mu_{y}}-\sqrt{\varepsilon_{x}\mu_{x}})\cos\theta_{i}}{({q_{e}}+\varepsilon_{x}\cos\theta_{i})({{q_{o}}+\mu_{x}\cos\theta_{i}})},
\label{Rco}
\end{eqnarray}
where $r_{pp}$, $r_{ss}$ and $r_{ps}(r_{sp})$ denote the Fresnel reflection coefficients for parallel, perpendicular and crossing polarizations, respectively. To verify the robustness of model, we set $\varepsilon_{x}=\varepsilon_{y}=const,\mathbf{\mu}=1$, the reflection coefficients Eqs. (\ref{Rco1}) and (\ref{Rco}) can reproduce to the Fresnel equation in the classic case of isotropic~\cite{BornWolf}.

\section{Giant Photonic Spin Hall effect}\label{SecII}

We now develop the theoretical mode to describe the behavior of Gaussian beam near the Dirac point at the interface. This model can be extensively extended to other 3D anisotropic photonics crystals. For horizontal polarization ($|{H}({k}_{i,r})\rangle$) and vertical polarization ($|{V}({k}_{i,r})\rangle$), the corresponding individual wave vector components can be expressed by $|{P}({k}_{i})\rangle$
 and $|{S}({k}_{i})\rangle$\cite{Hosten2008}:
\begin{eqnarray}
 |{H}({k}_{i,r})\rangle=|{P}({k}_{i,r})\rangle-\frac{k_{iy}}{k_{i,r}}\cot\theta_{i,r}|{S}({k}_{i,r})\rangle
\\
 |{V}({k}_{i,r})\rangle=|{S}({k}_{i,r})\rangle+\frac{k_{iy}}{k_{i,r}}\cot\theta_{i,r}|{P}({k}_{i,r})\rangle\label{VKIR}
\end{eqnarray}
 where $k_i$ and $k_r$ are the incident and reflected wave vectors respectively; $\theta_i$ is the incidence angle and $\theta_r$ is the reflection angle.

 The reflected angular spectrum of $|{P}({k}_{i})\rangle $ and $|{S}({k}_{i})\rangle $ is associated with the boundary distribution by means of the relation $[|{P}({k}_r)\rangle~|{S}({k}_r)\rangle]^T={M}_{R}[|{P}({k}_i)\rangle~|{S}({k}_i)\rangle]^T$, here $M_{R}$ can be expressed as\cite{LuoH2011}:
\begin{equation}
\left[                 
\begin{array}{cc}   
r_{pp}-\dfrac{k_{ry}(r_{ps}-r_{sp})\cot\theta_{i}}{k_{0}}&r_{ps}+\dfrac{k_{ry}(r_{pp}+r_{ss})\cot\theta_{i}}{k_{0}} \\
r_{sp}-\dfrac{k_{ry}(r_{pp}+r_{ss})\cot\theta_{i}}{k_{0}}&r_{ss}-\dfrac{k_{ry}(r_{ps}-r_{sp})\cot\theta_{i}}{k_{0}} \label{RM}
\end{array}
\right].
\end{equation}

In the above equation, the boundary conditions $k_{rx}=-k_{ix}$ and $k_{ry}=k_{iy}$ have been introduced. 

We then obtain
\begin{eqnarray}
 |{H}({k}_{i})\rangle&\rightarrow&   r_{pp}  |{H}({k}_{r})\rangle\nonumber\\
 &&+\left[r_{sp}-\frac{k_{ry}\cot\theta_{i}(r_{pp}+r_{ss})}{k_{0}}\right]|{V}({k}_{r})\rangle\label{HKI},
\end{eqnarray}
\begin{eqnarray}
 |{V}({k}_{i})\rangle&\rightarrow& \left[r_{ps}+\frac{k_{ry}\cot\theta_{i}(r_{pp}+r_{ss})}{k_{0}}\right] |{H}({k}_{r})\rangle\nonumber\\
 &&+ r_{ss}  |{V}({k}_{r})\rangle\label{VKI},
\end{eqnarray}
where $\theta_{i}$ is the incident angle and $k_0=\omega/c$ is the wave vector in vacuum.

The photonic SHE manifests itself as spin-dependent splitting which appears at transverse and longitudinal directions. Therefore, we now determine the spatial shifts of the wave packet. The polarization of $|{H}\rangle$ and $|{V}\rangle$ can be decomposed into two orthogonal spin components $|{H}\rangle=\frac{1}{\sqrt{2}}(|{+}\rangle+|{-}\rangle)$ and $|{V}\rangle=\frac{1}{\sqrt{2}}i(|\mathbf{-}\rangle-|{+}\rangle)$, where $|{+}\rangle$ and $ |{-}\rangle$ represent the left- and right-circular polarization components, respectively. And we assume that the wavefunction in momentum space can be specified by the following expression:
\begin{equation}
|\Phi\rangle=\frac{w_{0}}{\sqrt{2\pi}}\exp\left[-\frac{w^{2}_{0}(k_{ix}^{2}+k_{iy}^{2})}{4}\right]\label{GaussianWF},
\end{equation}
where $w_{0}$ is the width of wavefunction. From Eqs.~(\ref{RM})-(\ref{GaussianWF}) we get the corresponding reflected wavefunction in the momentum space, which are made up of the packet spatial extent and polarization description:
\begin{figure}
\centering
{\includegraphics[width=8.5cm]{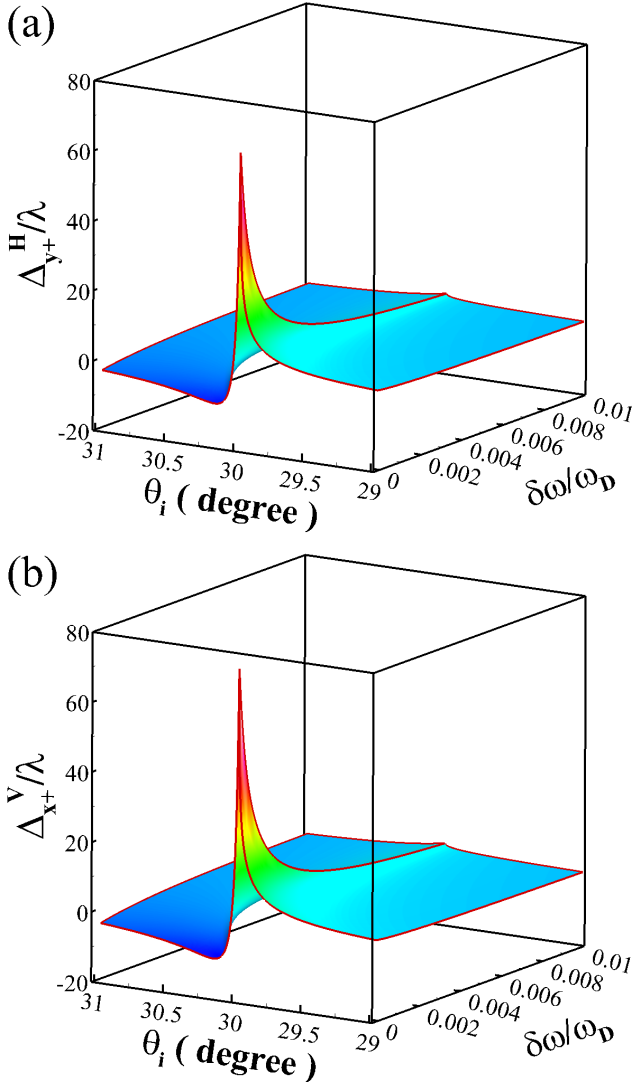}}
\caption{\label{Fig2}
The photonic SHE on the surface of Dirac metacrystal. The spatial shift of $|{H}\rangle$ polarization (a) and the spatial shift of $|{V}\rangle$ polarization (b) are plotted as the function of $\delta\theta_{i}$ and $\delta\omega$. $\delta\theta_{i}$ describe the change of incident angle near the critical angle $\theta_{c}=30^{\circ}$, $\delta\omega$ represent variable value of incident frequency $\omega$ relative to Dirac frequency $\omega_{D}$. We assume
an incident beam with waist radius $w_0$=100$\lambda$, and $\lambda$ =$2\pi c/\omega$, parameters for the Dirac metacrystal are chosen as $\varepsilon_{y}=\mu_{y}=0.5$, the resonance frequency $\omega_{0}=10^{15}Hz$ and structural coefficients $f_{1}$=0.5.
}
\end{figure}
\begin{figure}
\centering
{\includegraphics[width=8.6cm]{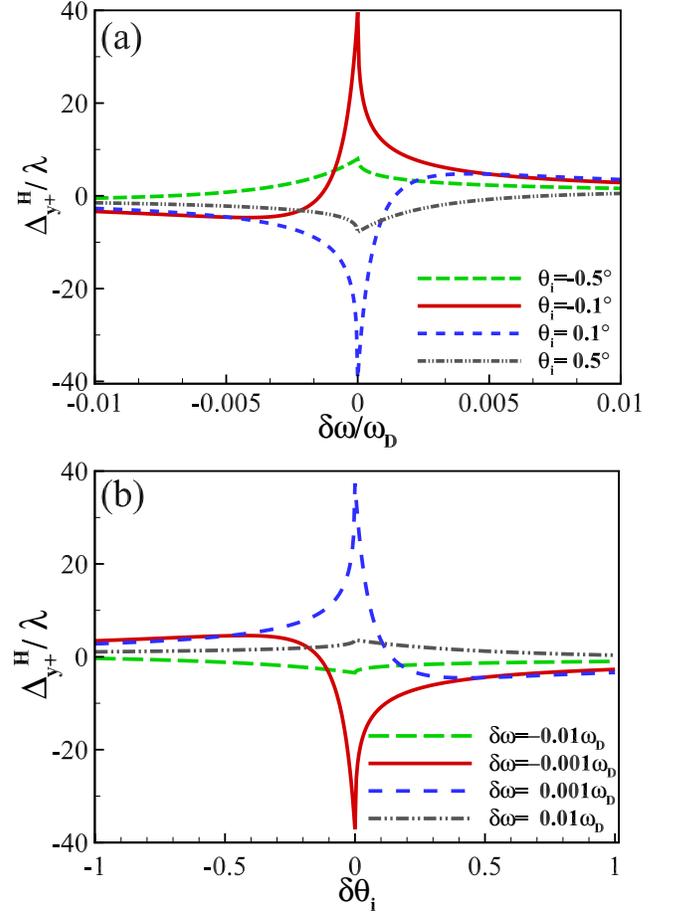}}
\caption{\label{Fig3}
The photonic SHE for $|{H}\rangle$  polarization with dfferent conditions (a) The transverse shift as a function of incident angle $\delta\theta_{i}=-0.5^{\circ}, -0.1^{\circ}, 0.1^{\circ}$ and $0.5^{\circ}$. (b) The transverse shift for incident frequency near Dirac frequency, we set the change $\delta\omega/\omega_{D}=-0.01, -0.001,  0.001, 0.01$, respectively. Other parameters are the same as those in Fig. 2.
}
\end{figure}
\begin{eqnarray}
|{\Phi}_r^{H}\rangle&\approx&\dfrac{1}{\sqrt{2}}\left\{r_{pp}-i { r_{sp}}+i k_{ry}\dfrac{(r_{pp}+r_{ss})\cot\theta_{i}}{k_{0}}\right\}|+\rangle|\Phi\rangle
    \nonumber \\
&&+\dfrac{1}{\sqrt{2}}\left\{r_{pp}+i { r_{sp}}-i k_{ry}\dfrac{(r_{pp}+r_{ss})\cot\theta_{i}}{k_{0}}\right\}|-\rangle|\Phi\rangle      \nonumber \\
&&=\frac{r_{pp}}{\sqrt{2}}\exp(+ik_{ry}\delta_{y+}^H)|+\rangle|\Phi\rangle   \nonumber \\
&&+\frac{r_{pp}}{\sqrt{2}}\exp(-ik_{ry}\delta_{y-}^H)|-\rangle|\Phi\rangle,   \label{WPVI}
\end{eqnarray}

\begin{eqnarray}
|{\Phi}_r^{V}\rangle&\approx&\dfrac{1}{\sqrt{2}}\left\{-i r_{ss}+ {r_{ps}}+ k_{ry}\dfrac{(r_{pp}+r_{ss})\cot\theta_{i}}{k_{0}}\right\}|+\rangle|\Phi\rangle
    \nonumber \\
&&+\dfrac{1}{\sqrt{2}}\left\{i r_{ss}+ {r_{ps}}+ k_{ry}\dfrac{(r_{pp}+r_{ss})\cot\theta_{i}}{k_{0}}\right\}|-\rangle|\Phi\rangle      \nonumber \\
&&=\frac{-i r_{ss}}{\sqrt{2}}\exp(+ik_{rx}\delta_{x+}^V+ik_{ry}\delta_{y+}^V)|+\rangle|\Phi\rangle   \nonumber \\
&&+\frac{i r_{ss}}{\sqrt{2}}\exp(-ik_{rx}\delta_{x-}^V-ik_{ry}\delta_{y-}^V)|-\rangle|\Phi\rangle.   \label{WPVI2}
\end{eqnarray}

Here we introduced the approximation of  $1\pm i k_{ry}\delta^{H,V}_{y}\approx\exp(\pm i k_{ry}\delta^{H,V}_{y})$ and $1\pm i k_{rx}\delta^{H,V}_{x}\approx\exp(\pm i k_{rx}\delta^{H,V}_{x})$. In addition, the higher order terms of $k_{rx}$ and $k_{ry}$ have been neglected, and only the first order are retained. The transverse spatial shift and in-plane shift can be written as
\begin{eqnarray}
&&\langle{\triangle _{y \pm}^{H}}\rangle=\dfrac{\langle\Phi_r^{H}|\partial_{k_{ry}}|\Phi_r^{H}\rangle}{\langle\Phi_r^{H}|\Phi_r^{H}\rangle},  \\
&&\langle{\triangle _{x\pm}^{V}}\rangle=\dfrac{\langle\Phi_r^{V}|\partial_{k_{rx}}|\Phi_r^{V}\rangle}{\langle\Phi_r^{V}|\Phi_r^{V}\rangle}.
\end{eqnarray}

Figure 2 shows the spatial shifts for the $|H\rangle$ and $|V\rangle$ polarization impinging on the surface of Dirac metacrystal. We set the critical incident angle as the $\theta_{c}=30^{\circ}(=\arcsin{\varepsilon_{y}\mu_{y}})$, $\delta\theta_{i}$ and $\delta\omega$ used to represent
the offset relative to Dirac point in the incident angle and frequency. Thus the spatial shifts are plotted as a function of $\delta\theta_{i}$ and $\delta\omega$. With the increase of incident angle reaching the critical angle, $\delta\theta_{i}=0$, the transverse and in-plane photonic SHE manifested as large spatial shifts occur near the Dirac point at the interface. Meanwhile, when the incident frequency meet the Dirac frequency $\omega_{D}$, the spatial shifts has grown more significantly, which corresponding to stronger spin-orbit interaction. Both the two polarization states have the similar behavior of beam, while the transverse shift is slightly larger than other,
So we choose the transverse photonic SHE as detailed discussions.

For certain incident angle, it can be seen that the spatial shifts are increased as the incident frequency approximating Dirac frequency $\omega_{D}$ (i.e.$\delta\omega=0$) as shown in Fig. 3(a), but it is worth noting that the quantized step widths can be significantly enhanced. Furthermore, if the Gaussian beam with the specific incident frequency impinging on interface, see in Fig. 3(b), the remarkable spin-orbit interaction exist in $\theta_{c}=30^{\circ}$, which can be considered as imaging near the Dirac point. And the opposite $\delta\omega$ ($\delta\theta_{i}$) bring equal and reversal spatial shifts. It should be interesting for the proposed metacrystals, the position of Dirac point can be determined by permittivity $\bm\varepsilon$ and permeability $\bm\mu$, thus we can enable a precise way to enhance the photonic SHE by constructing the certain Dirac points (or Weyl points, nodal degeneracy points), it also provide insights to measure the location of nodal degeneracy by photonic SHE in various photonic systems.

The photonic SHE can be described as a consequence of geometric phase which arises from the spin-orbit interaction.  We now examine the polarization rotation in the photonic SHE. The polarization states can be expressed in term of the Jones matrix~\cite{ZhouJX,MiPR}:
\begin{equation}
\left(
\begin{array}{c}   
\cos\nu  \\
\sin\nu
\end{array}
 \right)
 =\exp(+i\phi_{G})|+\rangle+\exp(-i\phi_{G})|-\rangle,
\end{equation}
where $\nu$ is the polarization angle. The polarization rotation will induce a geometric phase gradient  which can be regarded as the physical origin of photonic SHE. When the polarization rotation occurs in momentum space, the spatial shift in position space will be generated, which can be written as
\begin{eqnarray}
\langle{\triangle _{y \pm}^{H}}\rangle=\dfrac{\partial\phi_G}{\partial{k_{ry}}}=\sigma {\rm Re}[\delta_{y\pm}^H],
\label{sig}
\end{eqnarray}
where $\sigma=\pm1$. Based on the boundary conditions $k_{rx}=-k_{ix}$ and  $k_{ry}=k_{iy}$, the angular spectrum of reflected field can be written as

\begin{eqnarray}
|\psi_{r}^H\rangle&=&\exp\left[-\frac{w^{2}_{0}(k_{rx}^{2}+k_{ry}^{2})}{4}\right]\bigg[r_{pp}|{H}\rangle\nonumber\\
 &&+\left(r_{sp}-\dfrac{k_{ry}\cot\theta_{i}(r_{pp}+r_{ss})}{k_{0}}\right)|{V}\rangle\bigg],\label{HKIC}
\end{eqnarray}
\begin{eqnarray}
|\psi_{r}^V\rangle&=&\exp\left[-\frac{w^{2}_{0}(k_{rx}^{2}+k_{ry}^{2})}{4}\right]\bigg[r_{ss}|{V}\rangle\nonumber\\
 &&+\left(r_{ps}+\dfrac{k_{ry}\cot\theta_{i}(r_{pp}+r_{ss})}{k_{0}}\right)|{H}\rangle\bigg].\label{VKIC}
\end{eqnarray}

The wave function in position space is the Fourier transform of the wave function in momentum space:
\begin{equation}
|\Phi_{r}^{H,V}\rangle=\int\int{dk_{rx}dk_{ry}}|\psi_{r\pm}^{H,V}\rangle|k_{rx},k_{ry}\rangle.\label{Fourier}
\end{equation}

\begin{figure}
\centering
{\includegraphics[width=8.6cm]{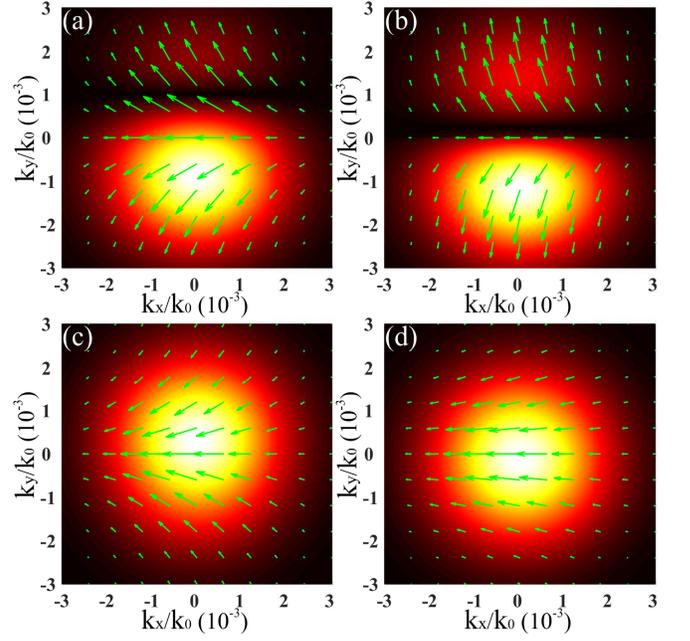}}
\caption{\label{Fig4} The role of incident frequency in polarization distributions of the $|{H}\rangle$ polarization reflected at the interface of Dirac metacrystal. (a) $\delta\omega=-0.01\omega_{D}$, (b) $\delta\omega=-0.001\omega_{D}$, (c) $\delta\omega=0.001\omega_{D}$, (d) $\delta\omega=0.01\omega_{D}$.
The incident angle is chosen as $\theta_{i}=30.1^{\circ}$ ($\delta\theta_{i}=0.1^{\circ}$) and other parameters are the same as in Fig. 2.
}
\end{figure}
\begin{figure}
\centering
{\includegraphics[width=8.6cm]{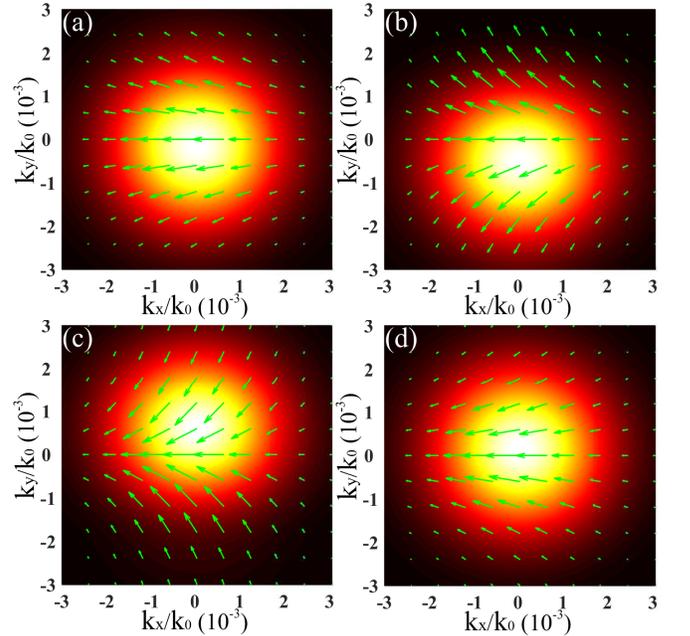}}
\caption{\label{Fig5}The role of incident angle in polarization distributions of the $|{H}\rangle$ polarization reflected at the interface of Dirac metacrystal. (a) $\delta\theta_{i}=-0.5^{\circ}$, (b) $\delta\theta_{i}=-0.1^{\circ}$, (c) $\delta\theta_{i}=0.1^{\circ}$, (d) $\delta\theta_{i}=0.5^{\circ}$.
The incident frequency is chosen as $\omega=(1+0.001)\omega_D$ and other parameters are the same as in Fig. 2.
}
\end{figure}

After substituting Eqs.~(\ref{HKIC}) and~(\ref{VKIC}) into Eq.~(\ref{Fourier}), the general representation of reflected wave function in position space can be written as:
\begin{eqnarray}
|\Phi_{r}^H\rangle&=&\exp\left[-\frac{(x_{r}^{2}+y_{r}^{2})}{w^{2}_{0}}\right]\bigg[r_{pp}|{H}\rangle\nonumber\\
 &&+\left(\dfrac{i y_{r}}{z_{R}} r_{sp}'-\dfrac{iy_r\cot\theta_{i}(r_{pp}+r_{ss})}{z_R} \right)|{V}\rangle\bigg]\label{HPR},
\end{eqnarray}
\begin{eqnarray}
|\Phi_{r}^V\rangle&=&\exp\left[-\frac{(x_{r}^{2}+y_{r}^{2})}{w^{2}_{0}}\right]\bigg[r_{ss}|{V}\rangle\nonumber\\
 &&+\left(\dfrac{i x_{r}}{z_{R}} r_{ps}'+\dfrac{iy_r\cot\theta_{i}(r_{pp}+r_{ss})}{z_R} \right)|{H}\rangle\bigg]\label{VPR},
\end{eqnarray}
where $r_{sp}=r_{sp}' \frac{k_{ry}}{k_{0}}$, $r_{ps}=-r_{ps}' \frac{k_{rx}}{k_{0}}$, and $z_R=k_0 w_0^2/2$ is the Rayleigh length.

We plot the polarization distributions of the $|{H}\rangle$ polarization state as shown in Fig. 4 and Fig. 5, respectively. Figure 4 shows the polarization distributions of the $|{H}\rangle$ polarization reflected with different incident frequency.
Let the incident angle as $\theta_{i}=30.1^{\circ}$, there is a obvious rotation for the incident frequency $\delta\omega=-0.01\omega_{D}$, as shown in Fig. 4(a). Then, we get a remarkable polarization rotation with $\delta\omega=-0.001\omega_{D}$ in Fig. 4(b). Moreover, when the deviation of frequency transform from $-0.001\omega_{D}$ to $0.001\omega_{D}$, we can observed significant transition of polarization rotation near the Dirac point, a clockwise rotation transform to a anticlockwise rotation, as shown in Fig. 4(c). Furthermore, Fig. 4(d) shows the slight polarization rotation when we increase incident frequency to $\delta\omega=0.01\omega_{D}$.

Compared with Fig. 4, Fig. 5 describes the polarization distributions for several incident angles with certain frequency $\omega=\omega_{D}+0.001\omega_{D}$. When we choose the incident condition as $\delta\theta_{i}=-0.5^{\circ}$, Fig. 5(a) shows a weak rotation distribution. Similarly, the small polarization rotation also occur with the case of $\delta\theta_{i}=0.5^{\circ}$, see in Fig. 5(d).
When the incident condition further approximating the critical angle of Dirac point, we can also observed significant and opposite polarization rotation rate with $\delta\theta_{i}=\pm 0.1^{\circ}$, as shown in Fig. 5(c) and Fig. 5(d). The above homologous geometric phase gradient unveiled the giant photonic SHE near the Dirac point. The origin of the spin-orbit interaction lies in the transverse nature of the photon polarization: The polarizations associated with the different plane-wave components experience different rotations in order to satisfy the transversality of photon polarization. The wave packet experiences inhomogeneous geometric phases, whose gradient will result in the spin-dependent splitting.

\section{Conclusions}
In summary, we have developed a general model to describe the photonic SHE in three dimesional Dirac metacrystal and revealed giant photonic spin Hall effect near the Dirac point. When the Gaussian beam impinges on the
unique interface, we found that the strong photonic SHE near the dirac point manifesting itself as large spin-dependent splitting in position spaces. The remarkable transversal and in-plane shifts upon the condition of Dirac point, including electromagnetic properties, incident angle and frequency. These findings provide a pathway for enhance the photonic SHE by constructing Dirac point and thereby open the possibility of developing various photonic devices. Our model can also be applied to describe the beam shifts in other 3D topological photonic systems due to
their similar topological structure. We believe that the investigation of giant photonic spin Hall effect near the Dirac point to be of fundamental significance and may provide a possible scheme for the measurement of Dirac points, even Weyl points and other degeneracy points.

\section*{ACKNOWLEDGMENTS}
This research was supported by the National Natural Science
Foundation of China (Grant Nos. 61835004).


\begin{references}

\bibitem{Jackson1999} J. D. Jackson, \emph{Classical Electrodynamics} (Wiley, New York,
1999).
\bibitem{Onoda2004}M. Onoda, S. Murakami, and N. Nagaosa, Phys. Rev. Lett. \textbf{93}, 083901 (2004).
\bibitem{Hosten2008}O. Hosten and P. Kwiat, Science \textbf{319}, 787 (2008).
\bibitem{Bliokh2015}K. Y. Bliokh, F. J. Rodriguez-Fortuno, F. Nori, and A. V.
Zayats, Nat. Photonics \textbf{9}, 796 (2015).

\bibitem{Bliokh201502}K. Y. Bliokh, D. Smirnova, and F. Nori, Science \textbf{348}, 1448 (2015)

\bibitem{yinxb}X. B. Yin, Z. L. Ye, J. Rho, Y. Wang, and X. Zhang, Science \textbf{339}, 1405 (2013).
\bibitem{Ling2017} X. Ling, X. Zhou, K. Huang, Y. Liu, C.-W. Qiu, H. Luo, and S. Wen, Rep.
Prog. Phys. \textbf{80}, 066401 (2017).
\bibitem{Korger2014} J. Korger, A. Aiello, V. Chille, P. Banzer, C. Wittmann, N. Lindlein, C. Marquardt, and G. Leuchs, Phys. Rev. Lett. \textbf{112}, 113902 (2014).
\bibitem{zhoux2012} X. Zhou, X. Li, H. Luo, and S. Wen, Appl. Phys. Lett. \textbf{101}, 251602 (2012).
\bibitem{chens2017} S. Chen, C. Mi, L. Cai, M. Liu, H. Luo, and S. Wen, Appl. Phys. Lett. \textbf{110}, 031105 (2017).
\bibitem{ChenS2018} S. Chen, C. Mi, W. Wu, W. Zhang, W. Shu, H. Luo, and
S. Wen, New J. Phys. \textbf{20} 103050 (2018).
\bibitem{zhouxpra2012} X. Zhou, Z. Xiao, H. Luo, and S. Wen, Phys. Rev. A \textbf{85}, 043809 (2012).
\bibitem{qiu2014} X. Qiu, X. Zhou, D. Hu, J. Du, F. Gao, Z. Zhang, and H. Luo, Appl. Phys. Lett. \textbf{105}, 131111 (2014).

\bibitem{M¨¦nard2010}J.-M. M\'{e}nard, A. E. Mattacchione, H. M. van Driel, C. Hautmann, and M. Betz, Phys. Rev. B \textbf{82}, 045303 (2010).

\bibitem{Kamp2016}W. J. M. Kort-Kamp, N. A. Sinitsyn, and D. A. R. Dalvit, Phys. Rev. B \textbf{93}, 081410(R)
(2016).

\bibitem{cai2017}L. Cai, M. Liu, S. Chen, Y. Liu, W. Shu, H. Luo, and S. Wen, Phys. Rev. A \textbf{95}, 013809 (2017).

\bibitem{ChenWJ2014}W. J. Chen, S. J. Jiang, X. D. Chen, B. Zhu, L. Zhou, J. W. Dong, and C. T. Chan, Nat. Commun. \textbf{5}, 5782 (2014).
\bibitem{HeC2016}C. He, C, X. C. Sun, X. P. Liu, M. H. Lu, Y.Chen, L. Feng, and Y. F. Chen, Proc. Natl Acad. Sci. USA \textbf{113}, 4924 (2016).

\bibitem{Slobo2017}A. Slobozhanyuk, S. H. Mousavi, X. Ni, D. Smirnova, Y. S.
Kivshar, and A. B. Khanikaev, Nat. Photonics \textbf{11}, 130
(2017).
\bibitem{XuSY2015} S.-Y. Xu, I. Belopolski, N. Alidoust, M. Neupane, G. Bian, C. Zhang, R. Sankar, G. Chang, Z. Yuan, C.-C. Lee, S.-M. Huang, H. Zheng, J. Ma, D. S. Sanchez, B. Wang, A. Bansil, F. Chou, P. P. Shibayev, H. Lin, S. Jia, and M. Z. Hasan,
Science \textbf{349}, 613 (2015).
\bibitem{YangB2017}B. Yang, Q. Guo, B. Tremain, L. E. Barr, W. Gao, H. Liu, and S. Zhang, Nat. Commun. \textbf{8}, 97 (2017).

\bibitem{XiaoM2017}Xiao, M. Lin, Q. and Fan, S, Phys. Rev. Lett. \textbf{117}, 057401 (2016).


\bibitem{LiuZK2014}Z. K. Liu, B. Zhou, Y. Zhang, Z. J. Wang, H. M. Weng, D. Prabhakaran, and Z. Hussain, Science \textbf{343}, 864¨C867 (2014).
\bibitem{SOL2016}A. Slobozhanyuk, S. H. Mousavi, X. Ni, D. Smirnova, Y.  S. Kivshar, and A. B. Khanikaev, Nat. Photon. \textbf{11}, 130  (2016).
\bibitem{GuoQ2017}Q. Guo, B. Yang, L. Xia, W. Gao, H. Liu, J. Chen, Y. Xiang, and S. Zhang, Phys. Rev. Lett. \textbf{119}, 213901 (2017).


\bibitem{Cast2009}A. H. Castro Neto, F. Guinea, N. M. R. Peres, K. S.Novoselov,
and A. K. Geim, Rev. Mod. Phys. \textbf{81}, 109 (2009).

\bibitem{Wu2017} W. Wu, S. Chen, C. Mi, W. Zhang, H. Luo, and S. Wen, Phys. Rev. A \textbf{96}, 043814 (2017).
\bibitem{Kamp2018} W. J. M. Kort-Kamp, F. J. Culchac, R. B. Capaz, and F. A. Pinheiro,
Phys. Rev. B \textbf{98}, 195431 (2018).

\bibitem{Wu2018} W. Wu, W. Zhang, S. Chen, X. Ling, W. Shu, H. Luo, S. Wen, X. Yin, Opt. Express \textbf{26}, 23705 (2018).

\bibitem{Lul2016}L. Lu, C. Fang, L. Fu, S. G. Johnson, J. D. Joannopoulos,
and M. Soljacic, Nat. Phys. \textbf{12}, 337 (2016).

\bibitem{Lekner1991}J. Lekner, J. Phys.: Condens. Matter \textbf{3}, 6121 (1991).

\bibitem{ChenRL2014}P. H. Chang, C. Y. Kuo, R. L. Chern. Opt. Express \textbf{22}, 25710-25721(2014).

\bibitem{BornWolf}M. Born, and F. Wolf, \textit{Principles of optics: electromagnetic theory of propagation, interference and diffraction of light} (Elsevier, 2013)

\bibitem{LuoH2011} H. Luo, X. Zhou, W. Shu, S. Wen, and D. Fan,
Phys. Rev. A \textbf{84}, 043806 (2011).




\bibitem{ZhouJX}J. Zhou, W Zhang, Y. Liu, Y. Ke, Y. Liu, H. Luo, and S. Wen. Sci. Rep \textbf{6}, 34276.
\bibitem{MiPR} C. Mi, S. Chen, W. Wu, W. Zhang, X. Zhou, X. Ling, W. Shu, H. Luo,
and S. Wen, Opt. Lett. \textbf{42}, 4135 (2017).



\end{references}
\end{document}